\documentclass[12pt, oneside]{article}

\usepackage[colorlinks=true, linkcolor=blue,  citecolor=black, urlcolor=black   ]{hyperref}

\usepackage{subcaption}
\usepackage{graphicx}
\usepackage{booktabs}

\usepackage{caption}

\usepackage[left]{lineno} 

\newcommand{\secnameref}[1]{%
  {\hypersetup{linkcolor=black}\nameref{#1}}%
}

\usepackage[letterpaper, margin=1in]{geometry}

\usepackage{multirow}
\usepackage{amsmath}
\usepackage{xcolor}
\usepackage{makecell} 
\usepackage{comment}
\usepackage{natbib}

\usepackage{titlesec}

\titleformat{\subsection}
  {\normalfont\bfseries\normalsize}
  {\thesubsection}
  {1em}
  {}

\newcommand{\figref}[1]{%
  Fig~\textcolor{blue}{\ref{#1}}%
}

\newcommand{\tabref}[1]{Table~\textcolor{blue}{\ref{#1}}}

\title{Early warning of Mpox outbreaks in U.S. jurisdictions using Lasso Vector Autoregression models with cross-jurisdictional lags}

\begin{document}

        \maketitle

\begin{center}
\textbf{Hannah Craddock}$^{1}$, \textbf{Joel O. Wertheim}$^{1}$, \textbf{Eliah Aronoff-Spencer}$^{1}$, \textbf{Mark Beatty}$^{2}$, \textbf{David Valentine}$^{1}$, \textbf{Rishi Graham}$^{1}$, \textbf{Jade C. Wang}$^{3}$, \textbf{Lior Rennert}$^{4}$, \textbf{Seema Shah}$^{2^{*}}$, \textbf{Ravi Goyal}$^{1^{*}}$, \textbf{Natasha K. Martin}$^{1^{*}}$

\vspace{0.3cm}

$^{1}$ Division of Infectious Diseases and Global Public Health, University of California San Diego, San Diego, CA, USA \\
$^{2}$ County of San Diego Health and Human Services Agency, San Diego, CA, USA \\
$^{3}$ Bioinformatics and Systems Biology Graduate Program, University of California San Diego, La Jolla, CA, USA \\
$^{4}$ Department of Public Health, College of Behavioral, Social, and Health Sciences, Clemson University, Clemson, SC, USA \\

\vspace{0.1cm}
\end{center}

$^{*}$ Joint senior authors

\vspace{0.5cm}


\newpage
\section*{Abstract}
Mpox is an orthopoxvirus that infects humans and animals and is transmitted primarily through close physical contact. The episodic and spatially heterogeneous dynamics of Mpox transmission underscores the need for timely, area-specific forecasts to support targeted public health responses in the U.S. We develop a Vector Autoregression model with Lasso regularization (VAR-Lasso) to generate rolling two-week-ahead forecasts of weekly Mpox cases for eight high-incidence U.S. jurisdictions using national surveillance data from the Centers for Disease Control and Prevention (CDC). The VAR-Lasso model identifies significant long-lag, cross-jurisdictional predictors. For a case study in San Diego County (SDC), these statistical predictors align with phylogenetic analysis that traces a 2023 cluster in SDC to an outbreak in Illinois six months earlier. As the need for public health action is often greatest when incidence is increasing, our performance evaluation focuses on positive-slope weighted error metrics. Forecast performance of the VAR-Lasso model is compared to a uni-variate Auto-Regressive (AR) Lasso model and a naïve moving-average estimate. The models are compared using slope-weighted Root Mean Squared Error (RMSE), slope-weighted Mean Absolute Error (MAE), and slope-weighted bias. Across all observations, the VAR-Lasso model reduces slope-weighted RMSE, MAE, and bias by 12\%, 7\%, and 66\% relative to the AR model, and by 16\%, 13\%, and 76\% relative to the naïve benchmark. Our findings highlight the value of sparse multivariate time-series models that leverage cross-jurisdictional case data for early forecasting of Mpox outbreaks. Such forecasting can aid health departments in proactively providing timely resources and messaging to mitigate the risks of a future outbreak. 

\newpage
\section{Introduction}
Mpox is a zoonotic, infectious disease caused by the Mpox virus (MPXV), a double-stranded DNA virus of the genus Orthopoxvirus with two clades (Clade I and Clade II) \citep{mitja2023monkeypox}.  In 2022 there was a large global outbreak of Mpox Clade IIb which continues to circulate to this day. The 2022 outbreak affected regions - including the United States (U.S.) and Europe - that had previously reported few or no cases. In this outbreak, Mpox incidence has been markedly higher among men who have sex with men (MSM) than in other populations \citep{laurenson2023description}. This differs from the typical epidemiology observed in sub-Saharan Africa and suggests that in high-income countries (HICs), Mpox may circulate primarily among individuals with frequent close physical contact and particularly amongst MSM \citep{brand2023role}. The multifaceted response to Mpox has included expanding access to vaccination and developing targeted messaging for MSM aimed at reducing their risk of infection \citep{delaney2022strategies}. A highly effective vaccine Jynneos is available \citep{pischel2024vaccine} and several jurisdictions launched vaccine campaigns early in the 2022 outbreak \citep{owens2023jynneos}. Although the major U.S. outbreak subsided by late 2022, smaller outbreaks and intermittent periods of elevated incidence have continued across multiple jurisdictions \citep{CDC2025}. More recently, the first Clade I cases associated with community transmission in the U.S. were reported in California in October 2025 \citep{CDPH2025, CDC2025}. These developments highlight the continued need for robust early warning systems to support timely detection and public health response. 

Forecasting models have become critical tools for outbreak detection and response. From a public health perspective, the ability to have an early warning signal and to anticipate case increases ahead of time is valuable. Such early warning forecasts enable public health departments to plan effectively, including the coordination of vaccination strategies and educational campaigns \citep{overton2023nowcasting}. Uni-variate auto-regressive (AR) time‑series models are a common choice for forecasting Mpox \citep{munir2024time, bakare2025time}. However, such models cannot account for interdependent trends across multiple jurisdictions. Vector auto-regression (VAR) models extend AR methods to multivariate settings (Stock and Watson, 2001) and can model dynamic cross-jurisdictional relationships. However, as the number of jurisdictions and lags increases, VAR models become parametrically dense. Regularization methods such as the Least Absolute Shrinkage and Selection Operator (Lasso) address this challenge by shrinking some coefficients to exactly zero, effectively performing variable selection \citep{tibshirani1996regression}. The VAR-Lasso framework thus enables both forecasting and the identification of influential cross-jurisdictional lagged predictors. 

Despite its potential, to our knowledge, no prior work has applied VAR-Lasso to Mpox forecasting. This study applies a VAR-Lasso model to forecast weekly Mpox case counts for eight high-incidence U.S. jurisdictions using data from the U.S. Centers for Disease Control and Prevention (CDC). We conduct a detailed case study for San Diego County (SDC), California - one of the eight high-incidence jurisdictions - supported by an ongoing collaboration with the County of San Diego Health and Human Services Agency (CSDHHSA). This collaboration is part of Insight Net, a CDC network of 13 outbreak analytics and modeling centers across the U.S., in which Resilient Shield at the University of California, San Diego partners with the CSDHHSA. The work presented here forms part of an effort to develop an early warning system for detecting increases in Mpox incidence; accordingly, SDC provides a key operational context for evaluating and validating the forecasting model. 

In forecast evaluation, certain epidemiological features may be more consequential than others and can therefore be weighted accordingly when assessing predictive performance \citep{tabataba2017framework, armstrong2001evaluating}. As public health action is often most critical during periods of rising incidence, we introduce slope-weighted error metrics that weight forecast errors by the magnitude of positive week-to-week growth, ensuring that periods of increasing transmission receive proportionally greater emphasis in model evaluation.

The main objective of this study is to improve the timeliness and accuracy of Mpox case forecasts during periods of increase for high-incidence U.S. jurisdictions by leveraging cross-jurisdictional dependencies in reported case data. To achieve this we; (1) Develop a VAR-Lasso forecasting model with cross-jurisdictional lags to generate two-week-ahead forecasts of weekly Mpox case counts across eight high-incidence U.S. jurisdictions; (2) Externally validate the model by conducting phylogenetic analyses comparing the model-identified lagged predictors to phylogenetic findings for SDC; (3) Evaluate model performance during periods of increase using positive-slope-weighted error metrics that reflect the heightened need for public health action when incidence is increasing; and (4) Benchmark the VAR-Lasso forecasts against an AR-Lasso model and a naïve, moving-average estimate. Collectively, these steps develop, validate, and evaluate a sparse multivariate time-series framework, supported by phylogenetic evidence, that enhances short-term Mpox forecasting by integrating cross-jurisdictional case information and capturing long-lag transmission signals.

\section{Results}
First, we provide high-level descriptive statistics of Mpox cases in the U.S., highlighting the eight high-incidence jurisdictions. Second, we assess the overall forecasting performance of the VAR-Lasso model across all forecasts for 2024 and compare the results to a uni-variate AR Lasso model and naïve moving-average estimator. Third, we conduct a case study for SDC, evaluating its forecasts in detail and validating model predictions against independent phylogenetic data. Fourth, we report forecasts for the remaining jurisdictions, summarizing performance. Finally, we present additional sensitivity analyses.

\subsection{Descriptive statistics - Mpox cases in the U.S. and high-incidence jurisdictions}
\begin{figure}[tbp]
\centering
\includegraphics[width=18cm, height=17cm, keepaspectratio]{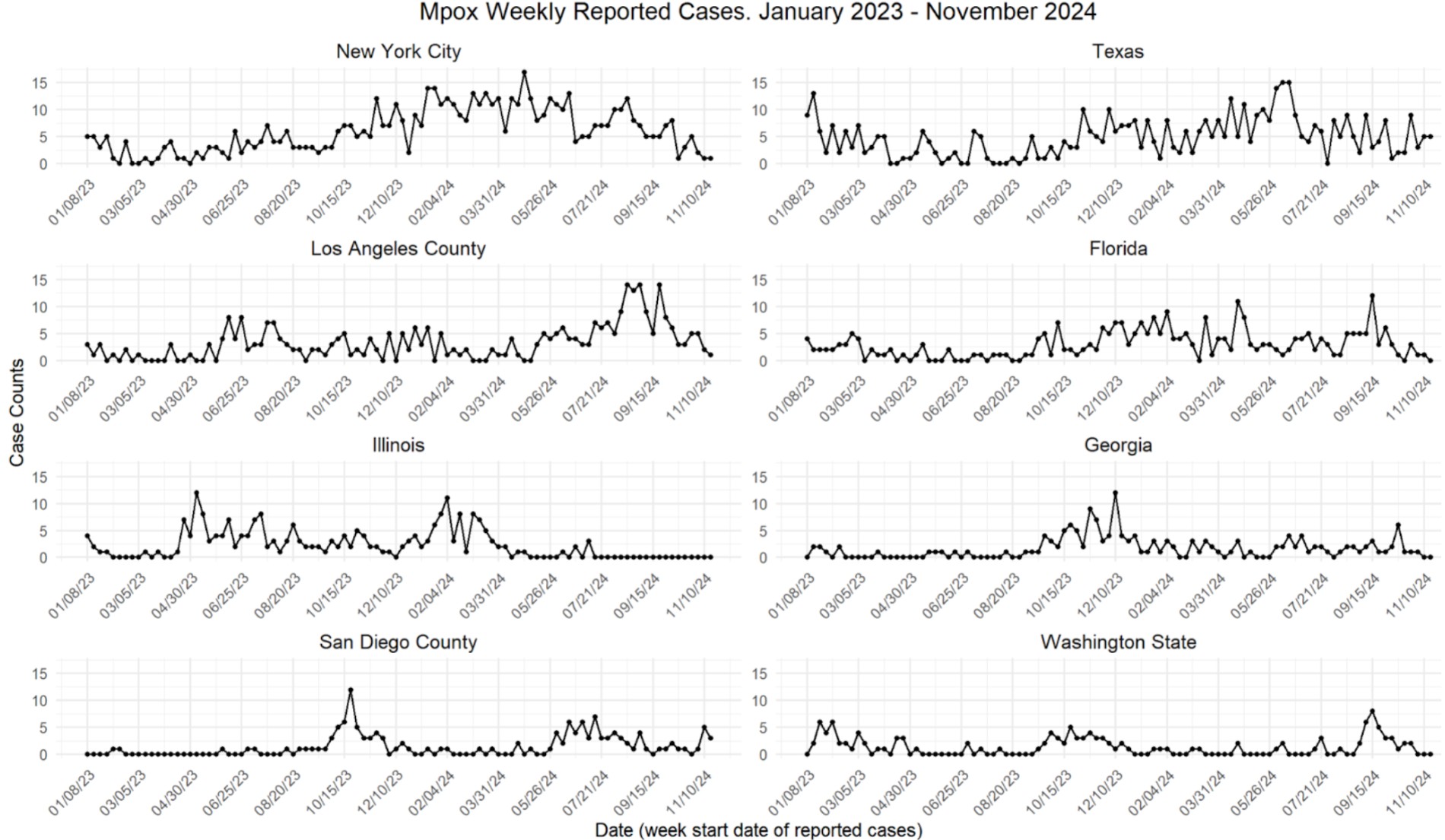}
\vspace{-0.5em} 
\caption{\textbf{Weekly reported Mpox cases for the 8 high-incidence U.S. jurisdictions from January 2023 - November 2024}. Obtained from the US Centers for Disease Control and Prevention (CDC) National Surveillance System.}
\label{fig:fig1}
\end{figure}

The large, global outbreak of Mpox in 2022 led to 30,069 reported cases in the U.S., spanning all 50 states. This was followed by 1,747 reported cases in 2023 and 2,529 reported cases from January to November 2024 (the most recent data from the CDC). We focus our modelling on the top eight high-incidence jurisdictions (which includes states, counties, and cities) that accounted for 55\% of all cases between 2022-2024 as summarized in \tabref{table:tab1}.  The eight jurisdictions are New York City, Texas (state), Los Angeles County, Florida (state), Illinois (state), Georgia (state), San Diego County, and Washington (state). We focus our modelling on the period commencing January 2023 until November 2024. The weekly reported cases for the top eight high-incidence jurisdictions during this period are displayed in \figref{fig:fig1}. 
\begin{table}[t]
\centering
\begin{tabular}{|l|c|c|c|}
\hline
\textbf{Jurisdiction} & \multicolumn{3}{|c|}{\textbf{Total yearly cases (\% of US yearly total)}} \\
\hline
 & \textbf{2022} & \textbf{2023} & \textbf{2024*} \\
\hline
New York City       & 3837 (13\%) & 198 (11\%) & 392 (16\%) \\
Texas               & 2969 (10\%) & 187 (10\%) & 283 (11\%) \\
Los Angeles County  & 2281 (8\%)  & 120 (7\%)  & 202 (8\%)  \\
Florida             & 2858 (10\%) & 109 (6\%)  & 178 (7\%)  \\
Illinois            & 1424 (5\%)  & 145 (8\%)  & 77 (3\%)   \\
Georgia             & 1986 (7\%)  & 90 (5\%)   & 70 (3\%)   \\
San Diego County    & 471 (1.5\%) & 59 (3\%)   & 77 (3\%)   \\
Washington state    & 653 (2\%)   & 81 (5\%)   & 47 (2\%)   \\
\hline
\textbf{Top 8 total cases} & \textbf{16,479 (55\%)} & \textbf{989 (57\%)} & \textbf{1326 (52\%)} \\
\textbf{US total cases}    & \textbf{30,069}        & \textbf{1,747}       & \textbf{2,529}       \\
\hline
\end{tabular}
\caption{\textbf{Yearly Mpox cases by jurisdiction and their corresponding percentage of total U.S. cases for selected high-incidence jurisdictions, 2022–2024}. Jurisdictions are ordered by total case counts in 2023–2024, the period included in the model. *Note the summary for 2024 is from January 2024 up to the week starting November 17$^{\text{th}}$ 2024 (the last available data from CDC).}
\label{table:tab1}
\end{table}

\subsection{VAR-Lasso forecast performance and benchmark model comparison}
\begin{figure}[tbp!]
\centering
\includegraphics[width=16cm, height=19cm, keepaspectratio]{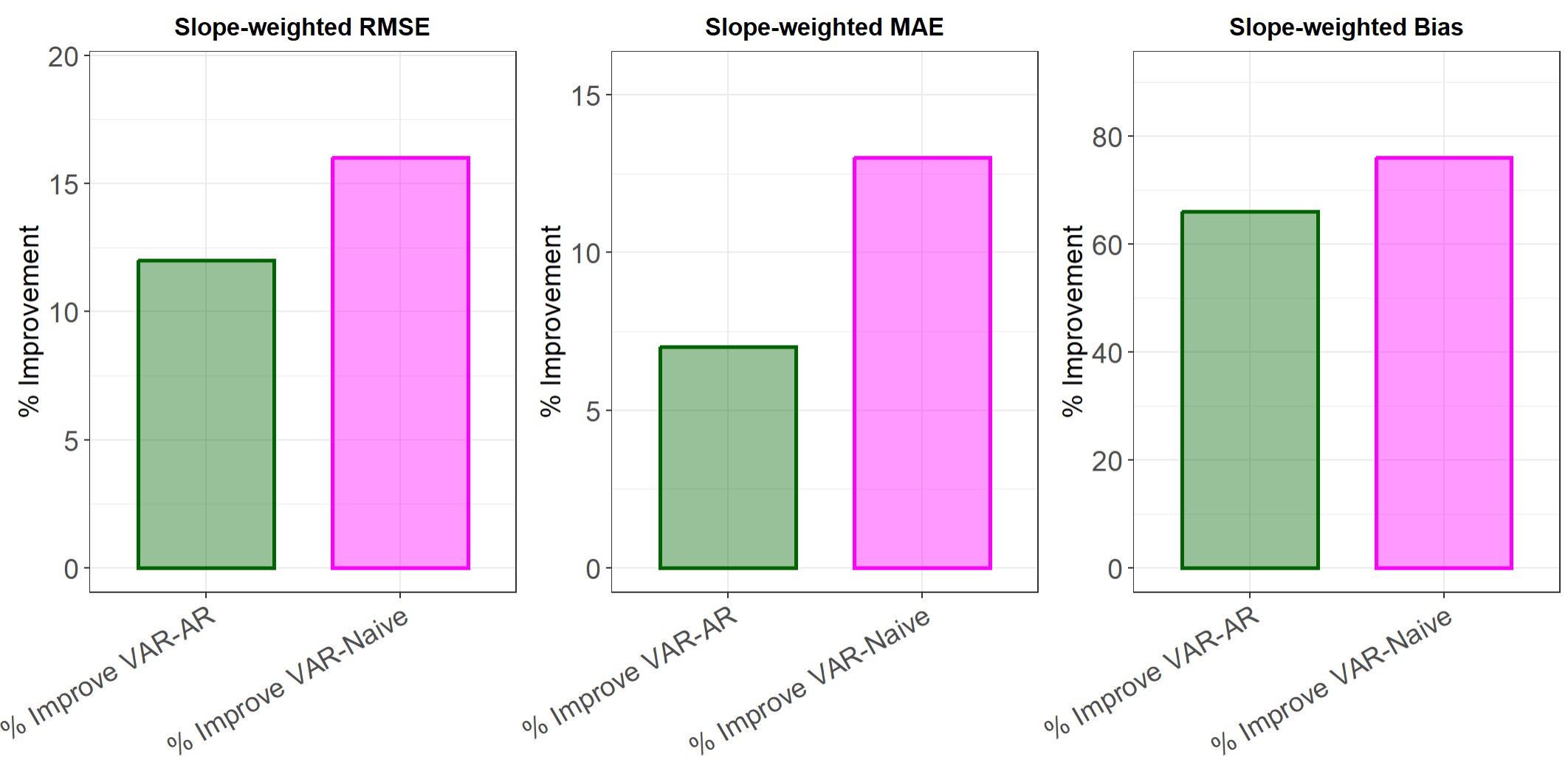}
\vspace{-0.5em} 
\caption{\small \textbf{ Percentage improvements of the VAR-Lasso model predictions compared to the AR-Lasso model predictions and the naïve moving-average estimate predictions based on slope-weighted metrics}. Slope-weighted RMSE, MAE, and bias are calculated for forecasts across all weekly predictions for the top eight high-incidence jurisdictions over 42 weeks from January to November 2024. Shown are the percentage improvements of the VAR-Lasso model predictions relative to the AR-Lasso predictions (\% Improve VAR-AR) and relative to the Naïve predictions (\% Improve VAR-Naive).}
\label{fig:fig2}
\end{figure}

We evaluate the predictive performance of the VAR-Lasso model for forecasting weekly Mpox cases for eight high-incidence U.S. jurisdictions in 2024, benchmarking it against a uni-variate AR-Lasso model and a naïve moving-average estimator (see \secnameref{sec:methods}). The model is initially trained on data from January–December 2023, with out-of-sample performance assessed on the 42 rolling, two-week-ahead forecasts in 2024. The training set is incrementally updated with each additional week of 2024 data. Across all forecasts, the VAR-Lasso consistently outperforms both benchmarks on all slope-weighted metrics, demonstrating superior accuracy in capturing increases in incidence (\figref{fig:fig2}; \tabref{table:tab2}). The VAR-Lasso model achieves a slope-weighted RMSE of 1.75, compared with 2.0 and 2.11 for the AR-Lasso and naïve benchmarks, corresponding to relative improvements of 12\% and 16\%, respectively. The slope-weighted RMSE of 1.75 for the VAR-Lasso model indicates that, when weighted by the observed positive weekly changes in cases, the model’s forecasts deviate from the observed trend by less than two cases per week across jurisdictions, capturing the magnitude of changes in incidence more accurately than the AR-Lasso and naïve benchmarks. Slope-weighted MAE is 1.4, versus 1.6 and 1.5 for the AR-Lasso and naïve models (7\% and 13\% improvement). The slope-weighted MAE of 1.42 for the VAR-Lasso model indicates that, when weighted by observed positive weekly changes in cases, forecasts deviate by about 1.4 cases per week, capturing the magnitude of weekly incidence changes reliably. Slope-weighted bias is reduced to –0.19 for the VAR-Lasso model, compared to –0.56 and –0.80 for the AR-Lasso and naïve benchmarks (66\% and 76\% improvement), indicating substantially lower systematic under-prediction relative to the other two models. 
\begin{table}[tbp]
\centering
\small 
\begin{tabular}{|l|c|c|c|c|c|}
\hline
\textbf{Model} & \multicolumn{3}{|c|}{\textbf{Evaluation Metric}} & \multicolumn{2}{|c|}{\textbf{\% Improvements of VAR}} \\
\hline
 & VAR-Lasso & AR-Lasso & Na\"ive estimate & \% Improve$_{\text{VA}}$ & \% Improve$_{\text{VN}}$ \\
\hline
Slope-weighted RMSE  & 1.75  & 2.00  & 2.11  & 12 & 16 \\
Slope-weighted MAE   & 1.42  & 1.52  & 1.64  & 7  & 13 \\
Slope-weighted Bias  & -0.19 & -0.56 & -0.80 & 66 & 76 \\
\hline
\end{tabular}
\caption{\small \textbf{Slope-weighed error metrics across all weekly forecasts for the top eight high-incidence jurisdictions over 42 weeks from January to November 2024}. The slope-weighted RMSE, slope-weighted MAE and slope-weighted bias are computed for the forecasts generated using the VAR-Lasso model, AR-Lasso model and the Naive moving-average estimator respectively. The \% improvement of the VAR-Lasso model predictions compared to the AR-Lasso model predictions (\% Improve$_{\text{VA}}$) and the \% improvement of the VAR-Lasso model predictions compared to the Naive estimate predictions (\% Improve$_{\text{VN}}$) are shown in the last two columns.}
\label{table:tab2}
\end{table}

\vspace{6mm}

\clearpage
\subsection{Case Study: San Diego County - forecast results \& phylogenetic validation}
SDC provides an operational setting for evaluating and validating the forecasting model, supported by the ongoing collaboration with CSDHHSA. SDC experienced low but recurrent Mpox transmission during 2023–2024, with intermittent periods of elevated incidence. This included a peak of twelve weekly cases in a single week in October 2023 and a period of increase in the summer of 2024, reaching a maximum of seven weekly cases that July. These fluctuations make SDC an informative setting for evaluating the model’s ability to anticipate short-term increases in cases. In line with the overall results, the VAR-Lasso model demonstrates improved performance for SDC relative to both benchmark models The VAR-Lasso model reduces slope-weighted RMSE, MAE, and bias by 11\%, 5\%, and 88\% relative to the AR model, and by 20\%, 17\%, and 90\% relative to the naïve benchmark (See \tabref{table:tab3}). 

\begin{figure}[tbp!]
\centering
\includegraphics[width=16.5cm, height=11cm, keepaspectratio]{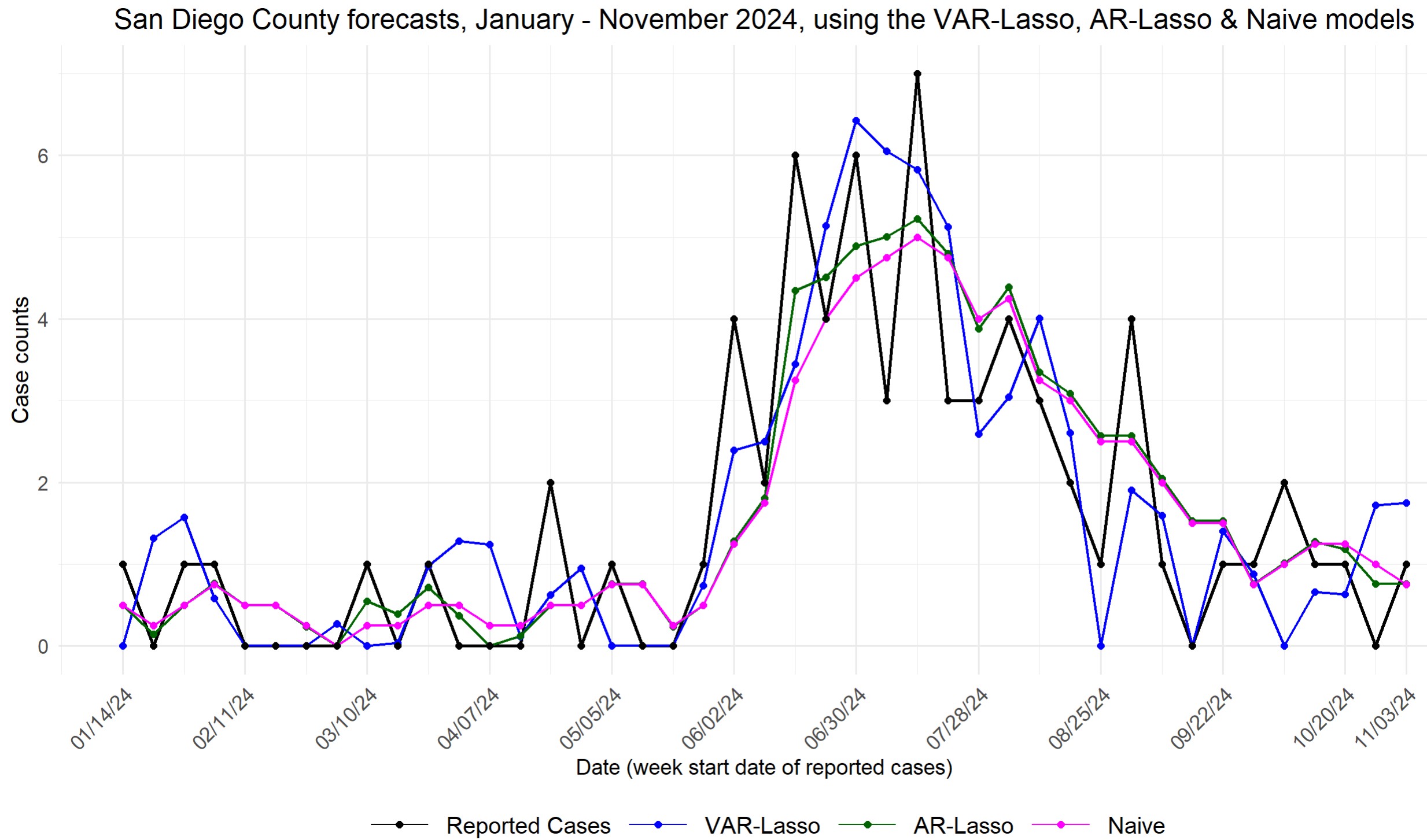}
\vspace{-0.5em} 
\caption{\textbf{Weekly reported Mpox cases and forecasts for San Diego County, January–November 2024}. Reported cases (black), VAR-Lasso forecasts (blue), AR-Lasso forecasts (magenta) and the naïve moving-average forecasts (green).}
\label{fig:fig3_forecasts_sd}
\end{figure}

\begin{figure}[tbp]
\centering
\begin{subfigure}{\linewidth}
    \centering
\includegraphics[scale = 0.6]{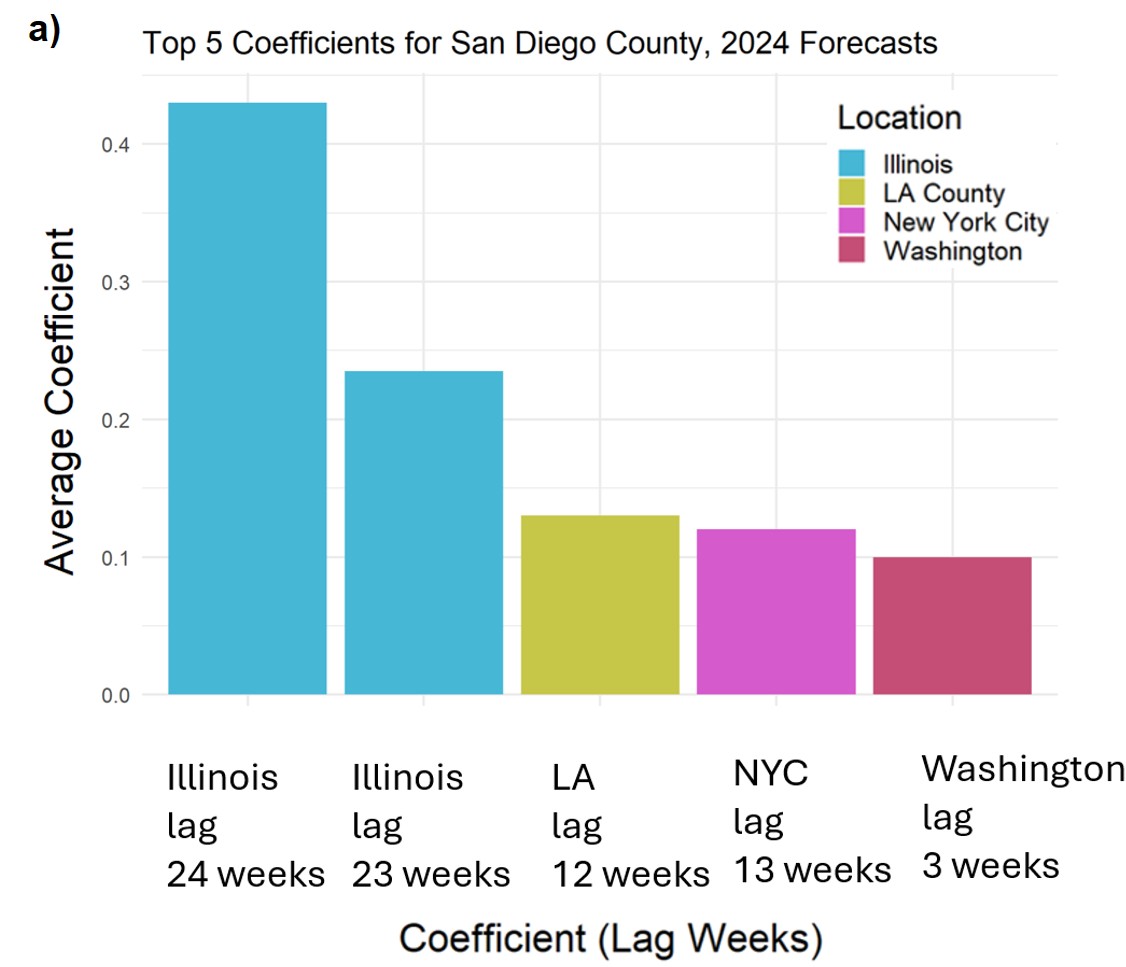}
\end{subfigure}
\begin{subfigure}{\linewidth}
    \centering
\includegraphics[scale = 0.62]{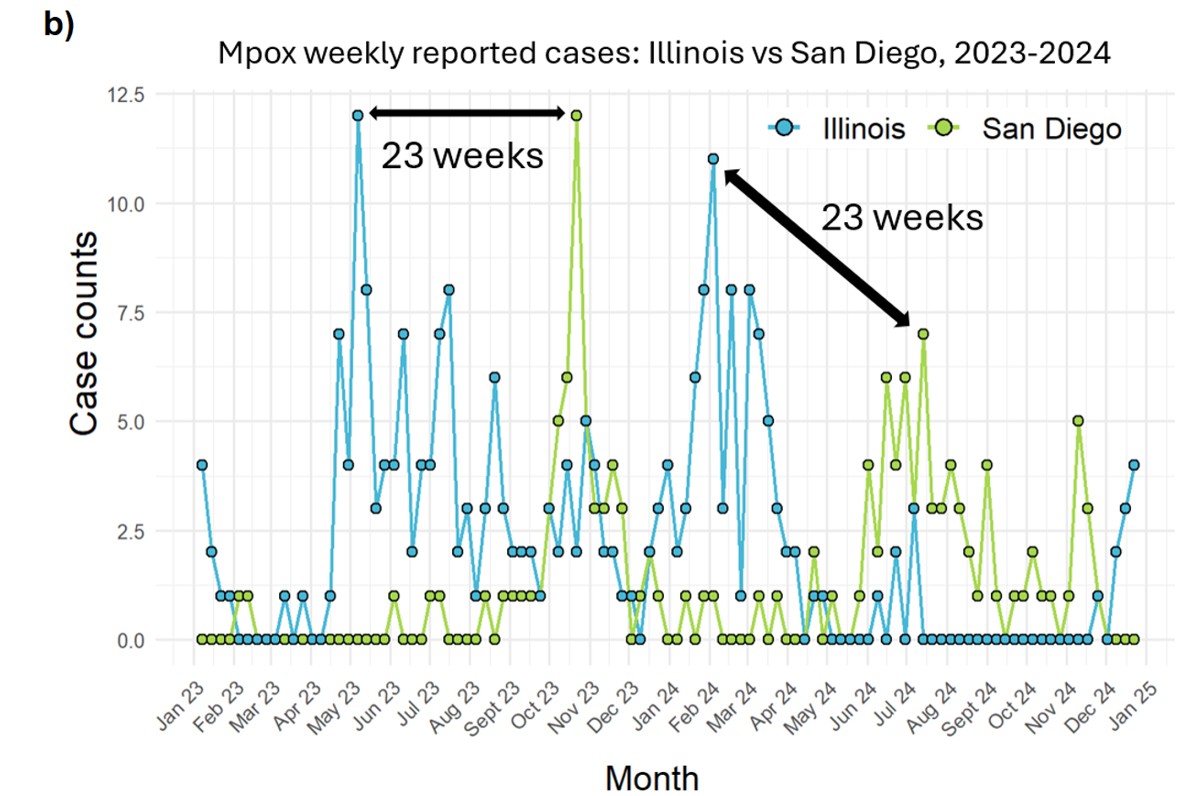}
\end{subfigure}
\caption{\small \textbf{The top five VAR Lasso model coefficients for San Diego County (SDC) and the weekly case counts for SDC and Illinois from 2023-24}. (\textbf{a}). Coefficients represent the mean magnitudes across all VAR–Lasso model fits used to forecast the 42 weeks in 2024. Each model is fit using all data from 2023 and data available up to two weeks prior to the forecasted week. The coefficients with the greatest magnitude correspond to Illinois at 23–24-week lags, followed by Los Angeles County (12-week lag), New York City (13-week lag), and Washington state (3-week lag). 
(\textbf{b}). Reported weekly Mpox cases for SDC and Illinois (2023–2024), showing Illinois peaks leading SDC by approximately 23 weeks, in agreement with phylogenetic evidence of introductions from Illinois.}
\label{fig:fig4}
\end{figure}

The weekly forecasts for SDC from January to November 2024 obtained using the VAR-Lasso, AR-Lasso, and the naïve estimator are shown in \figref{fig:fig3_forecasts_sd}. When cases begin rising in June 2024, peaking at seven weekly cases in July 2024, both the AR-Lasso and naïve estimator underestimate the surge. In contrast, the VAR-Lasso model can leverage cross-jurisdictional lagged case data to anticipate these peaks, producing forecasts that more closely match the observed maxima. The model identifies Illinois as the most influential external predictor for SDC, with the largest estimated coefficients corresponding to the 23- and 24-week lag terms (\figref{fig:fig4}a). The dominance of the 23–24 week lag terms implies a long-range cross-jurisdictional association, with Illinois case increases preceding SDC incidence by approximately six months. The six-month lag is evident in the raw case counts when Illinois and SDC reported weekly cases are overlaid (\figref{fig:fig4}b). In particular, Illinois experienced peaks in April 2023 and February 2024 that preceded peaks in SDC in October 2023 and July 2024 by approximately 23 weeks.
\\

\begin{figure}[tbp!]
\centering
\includegraphics[width=20cm, height=12cm, keepaspectratio]{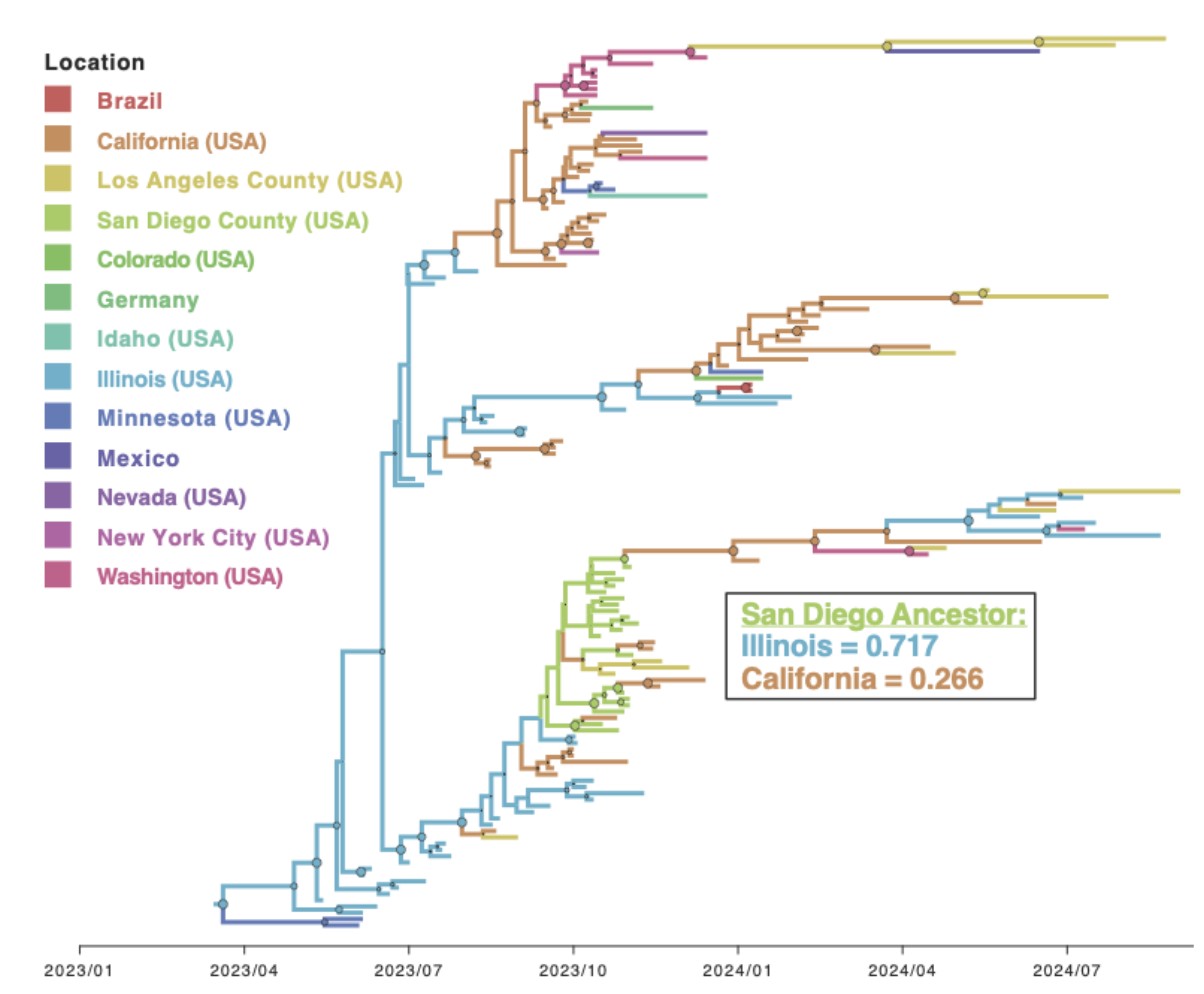}
\vspace{-0.5em} 
\caption{\textbf{Maximum clade credibility tree depicting migration of Mpox virus into San Diego County in late 2023}. Branches are colored based on inferred geographic location. Size of circles on internal nodes represent posterior support, and color represents geographic location. Insert denotes the posterior probability for the location of the branch immediately preceding the ancestor of the SDC MPXV genomes.}
\label{fig:fig5}
\end{figure}

\noindent \emph{External validation using phylogenetic analysis}. To validate the cross-jurisdictional and time-lagged predictors selected by the VAR-Lasso, we perform phylogeographic analysis to infer the timing and source of introductions of MPXV into SDC. Based on the 29 MPXV genomes associated with cases in SDC, we identify five distinct introductions during 2023–2024 (\tabref{table:tab4}). The largest cluster (Cluster B) includes 19 of the 29 viral genomes. The SDC sequences for Cluster B, sampled in October and November 2023, have a time of most recent common ancestor in September 2023 (\figref{fig:fig5}). Our phylogeographic inference indicates that MPXV was introduced either from Illinois (posterior probability = 0.72), California (posterior probability = 0.27), or Los Angeles County (posterior probability = 0.02). Regardless of the immediate ancestral location of this introduced virus, the ultimate source of this cluster is MPXV circulating in Illinois in April 2023 (posterior probability = 1.0). This indicates approximately a 22–28 week time lag from Illinois to SDC.

This independent genomic evidence aligns closely with the VAR-Lasso results, which assign the largest effects to the 23–24 week lag terms (\figref{fig:fig4}a). Together, the statistical and phylogenetic analyses converge on the same inter-jurisdictional transmission pathway, linking Mpox activity in Illinois to subsequent peaks in SDC occurring 23–24 weeks later.

\clearpage
\subsection{Jurisdictional-level forecast performance across high-incidence jurisdictions}
\begin{figure}[tbp]
\centering
\begin{subfigure}{\linewidth}
    \centering
\includegraphics[scale=0.40]{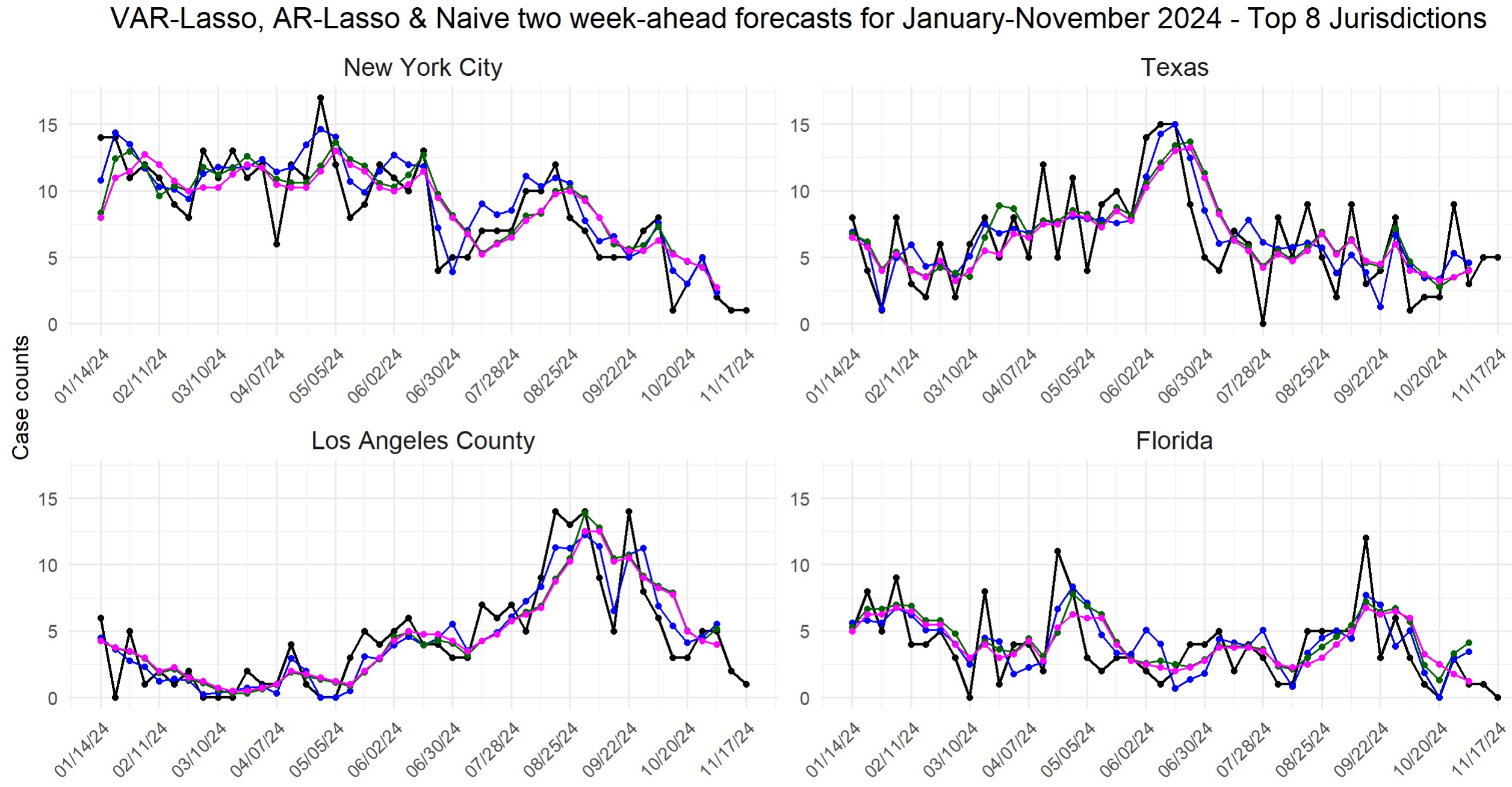}
\end{subfigure}
\vspace{0.2cm}
\begin{subfigure}{\linewidth}
    \centering
\includegraphics[scale=0.40]{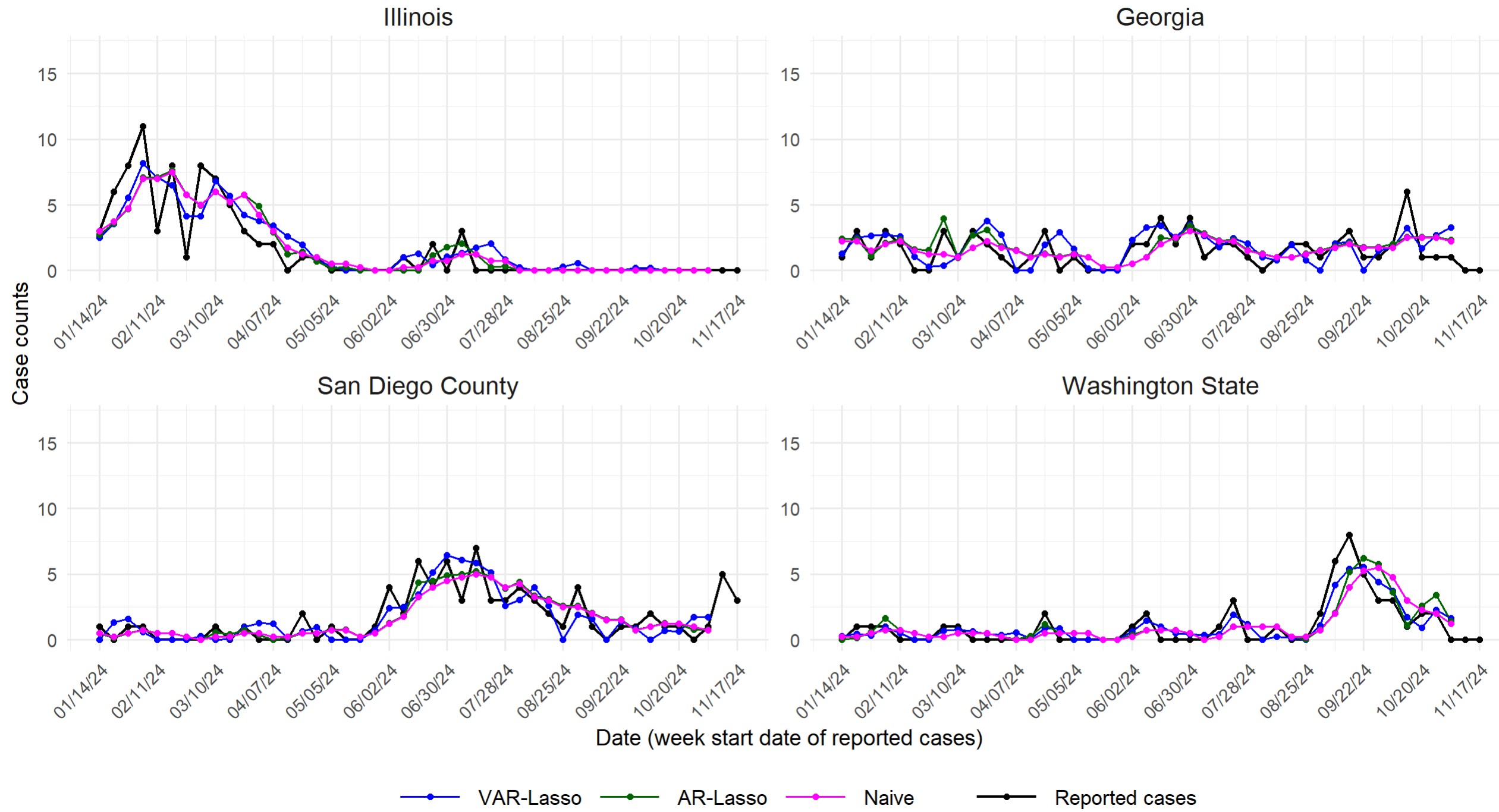}
\end{subfigure}
\caption{\textbf{Weekly reported Mpox cases and forecasts for the eight high-incidence US jurisdictions from January-November 2024}. Reported cases (black) and the forecasts generated using VAR-Lasso (blue), AR-Lasso (green), and the naïve estimator (magenta). See Supplementary section for the forecasts with prediction intervals.}
\label{fig:fig6_jur_forecats}
\end{figure}

We assess model performance at the jurisdictional level by calculating slope-weighted error metrics for each jurisdiction across their 42 weekly forecasts in 2024 (\figref{fig:fig6_jur_forecats}). The VAR-Lasso model outperforms the AR-Lasso model for 6 of the 8 jurisdictions in terms of slope-weighted RMSE and MAE, and outperforms the naïve benchmark in 7 of 8 jurisdictions (\tabref{table:tab3}). Improvements are observed for New York City, Texas, Los Angeles County, Illinois, San Diego County, and Washington. When an improvement is observed, slope-weighted RMSE reductions range from 4\% to 44\% relative to the AR-Lasso model and 7\% to 46\% relative to the naïve benchmark. Slope-weighted MAE reductions follow similar patterns, ranging from 5\% to 35\% versus the AR-Lasso model and 3\% to 40\% versus the naïve benchmark. Only Florida shows negative improvements in both RMSE and MAE. Overall, these results demonstrate that the VAR-Lasso model provides gains at the jurisdictional level. 

\begin{table}[tbp]
\centering
\small 
\begin{tabular}{|l|c|c|c|c|c|c|c|c|c|c|}
\hline
\textbf{Jurisdiction} & \multicolumn{5}{|c|}{\textbf{Slope-weighted RMSE}} & \multicolumn{5}{|c|}{\textbf{Slope-weighted MAE}} \\
\hline
 & VAR & AR & Naive & \% Imp$_{\text{VA}}$ & \% Imp$_{\text{VN}}$ & VAR & AR & Naive & \% Imp$_{\text{VA}}$ & \% Imp$_{\text{VN}}$ \\
\hline
New York City      & 1.50 & 1.60 & 2.10 & 6  & 26 & 1.3 & 1.3 & 1.7 & 0  & 25 \\
Texas              & 2.10 & 2.20 & 2.30 & 4  & 7  & 1.8 & 1.8 & 1.9 & 0  & 3  \\
L.A. County        & 1.95 & 2.60 & 2.70 & 25 & 26 & 1.7 & 2.2 & 2.2 & 24 & 24 \\
Florida            & 1.80 & 1.60 & 1.70 & -14 & -11 & 1.5 & 1.3 & 1.3 & -13 & -13 \\
Illinois           & 2.10 & 2.60 & 2.60 & 19 & 18 & 1.8 & 2.1 & 2.0 & 17 & 14 \\
Georgia            & 1.10 & 1.00 & 1.20 & -8 & 8 & 0.7 & 0.9 & 1.0 & 13 & 25 \\
San Diego Co.   & 1.20 & 1.35 & 1.50 & 11 & 20 & 1.0 & 1.1 & 1.2 & 5  & 17 \\
Washington & 1.00 & 1.80 & 1.80 & 44 & 46 & 0.7 & 1.1 & 1.1 & 35 & 40 \\
\hline
\end{tabular}
\caption{\textbf{Jurisdiction-level slope-weighted error metrics for weekly forecasts over 42 weeks, January–November 2024}. Columns include the slope-weighted RMSE and MAE for the forecasts obtained using VAR-Lasso, AR-Lasso the and naïve moving-average estimator as well as the  percentage improvement of the VAR-Lasso model relative to the AR-Lasso model (\% Imp$_{\text{VA}}$) and relative to the naïve estimator (\% Imp$_{\text{VN}}$).}
\label{table:tab3}
\end{table}

\newpage
\subsection{Sensitivity analysis }
\begin{figure}[tbp]
\centering
\begin{subfigure}{\linewidth}
    \centering
\includegraphics[scale = 0.6]
{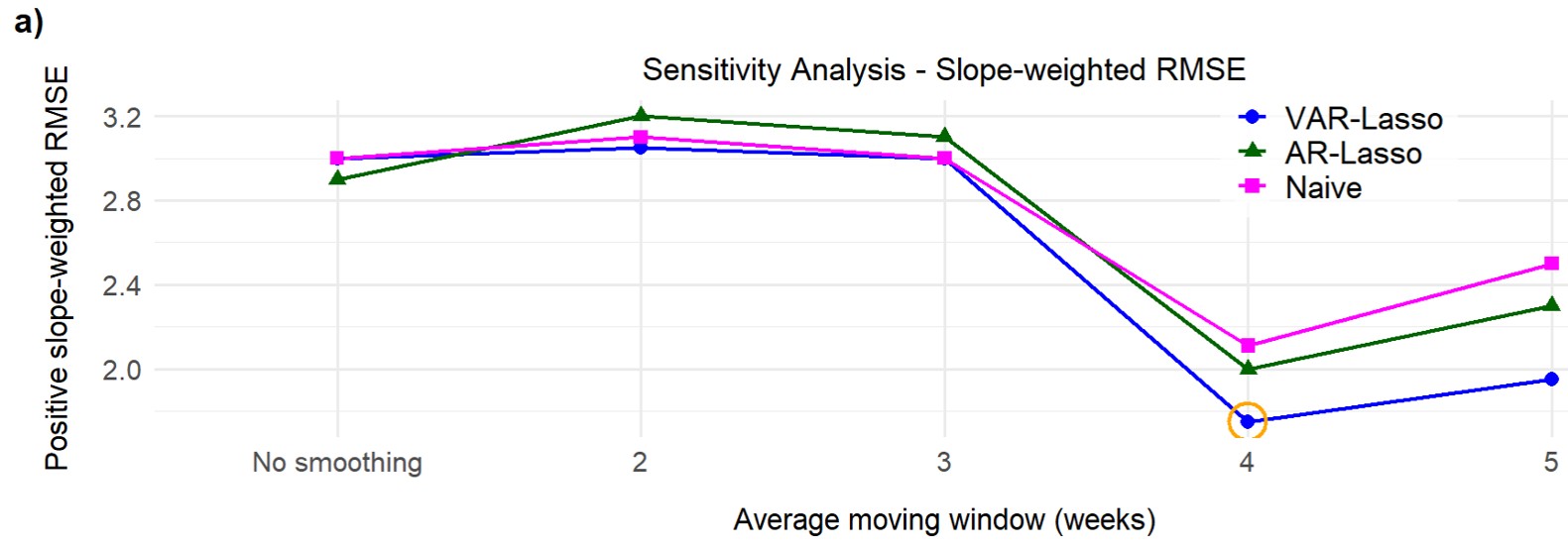}
\end{subfigure}
\vspace{1cm}
\begin{subfigure}{\linewidth}
    \centering
\includegraphics[scale = 0.6]{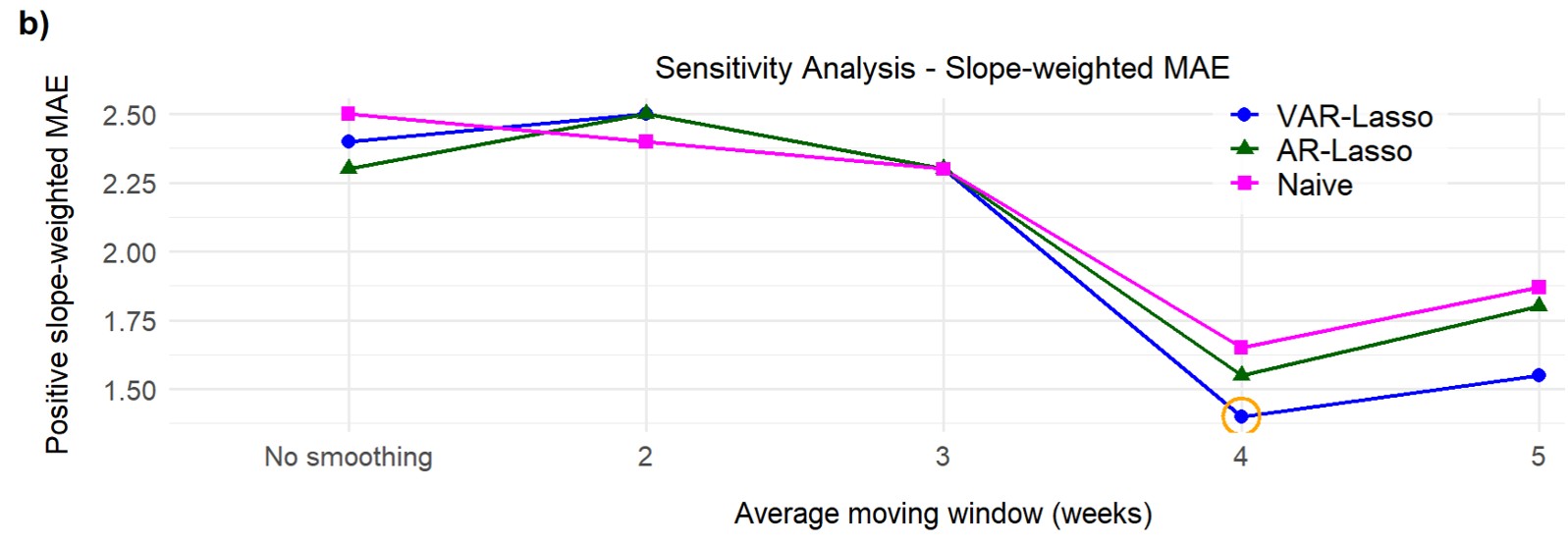}
\end{subfigure}
\caption{\textbf{Sensitivity analysis of the smoothing window duration on forecast performance measured by positive-slope-weighted RMSE (a) and positive-slope-weighted MAE (b)}. A sensitivity analysis is conducted to assess the impact of different smoothing windows (no smoothing, 2-, 3-, 4-, and 5-week windows) on forecast performance across the three models compared. The predictive model with the lowest error - the VAR-Lasso with a 4-week moving-average window - is highlighted with an orange circle.}
\label{fig:fig7}
\end{figure}

Weekly Mpox case counts exhibit stochastic noise (\figref{fig:fig1}). Smoothing may reduce the impact of these high-frequency fluctuations and improve the signal-to-noise ratio needed for analysis. We conduct a sensitivity analysis to assess the effect of smoothing window length on weekly Mpox data and subsequent forecast performance. To determine the optimal window length, the input time series are smoothed using centered moving averages of varying lengths (no smoothing, 2-, 3-, 4-, and 5-week windows). Forecasts are generated using each smoothing window for the VAR-Lasso, AR-Lasso, and naïve moving-average models, and performance is compared using positive slope-weighted error metrics. The slope-weighted RMSE and MAE across the 42 forecasted weeks for all eight jurisdictions, generated using different smoothing lengths, are shown in \figref{fig:fig7}. 

The results indicate that the choice of smoothing window substantially affects the forecast accuracy. Forecasts based on raw data or a 2-week window produce the largest errors, likely reflecting the strong influence of noise in the case counts on forecastability. Errors decrease when a 3-week window is used and reach a minimum using a 4-week moving average, which yields the lowest RMSE and MAE. However, performance declines again with further smoothing (5-week window), suggesting that excessive averaging obscures short-term epidemic dynamics.  Overall, the VAR-Lasso model with a 4-week smoothing window emerges as the optimal pre-processing choice, yielding the lowest errors. This smoothing configuration effectively balances noise reduction with the preservation of the epidemic signal, and is therefore used in the final predictive model.

\clearpage

\section{Discussion}
We have developed a novel VAR-Lasso Mpox forecasting model to generate two-week-ahead predictions of weekly Mpox case counts across eight high-incidence U.S. jurisdictions. The VAR-Lasso model achieves a superior predictive performance compared to the widely used AR approach and a naïve moving average baseline. 

This study presents a new application of a regularized multivariate time-series approach for infectious disease forecasting. Vector autoregression models provide a natural framework for capturing the dynamic interdependencies between jurisdictions, but they are prone to over-parameterization when the number of locations and lag terms grows, particularly in relatively short epidemic time series. Incorporating Lasso regularization addresses this challenge by shrinking many coefficients to zero, thereby reducing overfitting while automatically selecting informative lagged predictors. This produces a sparse and interpretable model structure that not only improves forecast performance but also highlights key cross-jurisdictional transmission pathways. 

The superior performance of the VAR-Lasso model over the more standard AR-Lasso and naïve approaches demonstrates the value of incorporating cross-jurisdictional dependencies. Many prior Mpox forecasting studies have relied only on univariate time-series approaches, such as autoregressive (AR) and ARIMA models, which capture temporal autocorrelation within a single jurisdiction but cannot incorporate cross-jurisdictional interactions. For example, Munir et al. (2024) used ARIMA models to forecast Mpox cases during the 2022 global outbreak across 10 heavily affected countries, and Bakare et al. (2025) applied a similar framework to produce long-term forecasts of Mpox cases and mortality in Nigeria. While these approaches can provide reasonable predictions within individual locations, they are inherently limited in their ability to detect or anticipate transmission driven by importations or geographically coupled dynamics. By contrast, the VAR-Lasso framework used in this study leverages multivariate dependencies across jurisdictions, allowing it to incorporate upstream case trends and identify informative cross-lagged predictors. This represents a methodological extension beyond existing Mpox forecasting work and highlights the added value of multivariate regularized models for settings where regional connectivity plays a substantive role in transmission patterns.

The agreement between the VAR-Lasso model’s identified lagged predictors and independent phylogenetic evidence represents a critical external validation of our modeling approach. By linking statistically inferred cross-jurisdictional effects from the VAR-Lasso model with actual genetic introductions of MPXV into SDC, in particular, the long 23–24 week lag from Illinois that mirrors the phylogenetic reconstruction of the dominant SDC cluster, we demonstrate that sparse multivariate models can identify upstream sources of infection before local outbreaks peak. These models can recover cross-jurisdictional dependencies that are independently supported by genomic epidemiology, highlighting their potential to provide an early statistical signal of inter-jurisdictional transmission dynamics of Mpox.

A further contribution of this work is the ability of the VAR-Lasso model to generate an early warning signal - two weeks ahead - of rising cases. From a public health perspective, the capacity to anticipate case increases ahead of time is valuable. Health officials at CSDHHSA have highlighted that such early warning forecasts can help to guide the timing of targeted messaging, outreach, and vaccination campaigns, enabling more efficient allocation of limited resources. 

Forecast evaluation metrics typically treat all time points equally, regardless of whether the underlying process is stable or rapidly changing. However certain epidemiological features are more influential than
others and may be weighted accordingly when evaluating predictive performance \citep{tabataba2017framework, armstrong2001evaluating}. The urgency for public health action is often greatest during periods of rising incidence, when vaccination and education campaigns are most critical. The use of slope-weighted error metrics address this point by prioritizing forecast performance during such periods of increasing transmission. 

This study has several important limitations. First, given the recent emergence of Mpox in the U.S., limited historical data restricts the model’s ability to learn longer-term seasonal or behavioral patterns, and future work could benefit from additional years of surveillance data. Second, under-diagnosis and under-reporting remain concerns for Mpox and may vary systematically across jurisdictions, potentially biasing both the case counts and the inferred cross-jurisdictional relationships. Third, the model relies solely on reported case data and does not incorporate other potentially informative signals - such as mobility patterns, sexual network data, vaccination coverage, or online data - which could enhance predictive performance and improve the interpretation of cross-jurisdictional dependencies. Fourth, while phylogenetic data provides valuable, external validation, the patient-level genomic sequencing data is only available for San Diego County. Finally, while the VAR-Lasso model outperforms both benchmarks across most observations and jurisdictions, it underperforms in Florida, where the AR-Lasso and naïve models yield more accurate forecasts. This discrepancy may reflect that incidence in certain jurisdictions is more strongly driven by local transmission. In such settings, uni-variate models that rely solely on a jurisdiction’s own historical data may be sufficient. 

Despite these limitations, our findings demonstrate that sparse, multivariate time-series models integrating cross-jurisdictional case information can substantially strengthen early detection and forecasting of Mpox activity in the U.S. The VAR-Lasso model’s ability to identify long-lag predictors consistent with phylogenetic evidence highlights its potential to anticipate outbreaks driven by geographically distant introductions. A promising future direction is the integration of novel data sources that could improve forecast accuracy and further support public health partnerships such as Resilient Shield and CSDHHSA in San Diego County. By providing actionable lead time for public health interventions, this approach illustrates the value of combining multivariate statistical modeling with epidemiological and genomic data to improve targeted public health action.

\section{Methods}
\label{sec:methods}

\subsection{Data}

Reported weekly Mpox case counts from January 1st, 2023 to November 24th, 2024 (the last available data) were obtained from the CDC's National Surveillance System. We select the eight U.S. jurisdictions with the highest cumulative Mpox incidence in 2023 and 2024, which together account for approximately 55\% of all reported U.S. cases across the two years. The eight selected jurisdictions, in order of total case counts in the period 2023–2024 are; New York City (city), Texas (state), Los Angeles County (county), Florida (state), Illinois (state), Georgia (state), San Diego County (county), and Washington (state). 

\subsection{Descriptive Statistics - Data}
Total and weekly cases are tabulated and visualized to characterize temporal trends and the relative contribution of each jurisdiction to national incidence. Summary statistics reported include the total cases per year and the proportion of U.S. cases represented by each jurisdiction.

\subsection{Time-series pre-processing: smoothing and differencing}
To reduce high-frequency fluctuations and improve the signal-to-noise ratio, weekly case counts are smoothed using a centered 4-week moving average. For each jurisdiction, the smoothed series replaces the original series in model fitting. A sensitivity analysis is conducted to assess the impact of different smoothing windows (2-, 3-, 4-, and 5-week windows). Each jurisdiction’s case time series was also tested for stationarity using the Augmented Dickey–Fuller (ADF) test \citep{dickey1979distribution}, as stationarity is required for fitting an autoregressive model. To achieve stationarity, each time-series is differenced once.

\subsection{The Vector Autoregression (VAR) Model}

To model the temporal dynamics and cross-jurisdictional dependencies in weekly Mpox case counts, we fit a VAR model to the processed time series as described above. Let $K$ denote the number of time series modeled; in our setting $K=8$, corresponding to the eight U.S. jurisdictions under consideration. Let $t \in \{1, 2, \dots\}$ index the weeks in our time series. For each week $t$, we define the $K$-dimensional vector $\mathbf{y}_t$ representing the observed weekly case counts as
\[
\mathbf{y}_t = (y_{1t}, y_{2t}, \dots, y_{Kt})^\top,
\]
where $y_{kt}$ represents the number of Mpox cases in jurisdiction $k$ during week $t$, and $\top$ denotes the transpose of the vector. The complete, observed time series can then be written as the sequence
\[
\mathbf{Y} = \{\mathbf{y}_1, \mathbf{y}_2, \dots\}.
\]

\noindent Let $\widetilde{y_{kt}}$, $\widetilde{\mathbf{y}_t}$ , and $\widetilde{\mathbf{Y}}$ denote the smoothed data. Our VAR model of lag order $p$ expresses each week’s smoothed data $\widetilde{\mathbf{y}_t}$ as a linear function of the $p$ most recent previous weekly observations, along with an intercept and a random error term:

\[
\widetilde{\mathbf{y}_t} = \mathbf{c} + \Phi_1 \widetilde{\mathbf{y}_{t-1}} + \Phi_2 \widetilde{\mathbf{y}_{t-2}} + \cdots + \Phi_p \widetilde{\mathbf{y}_{t-p}} + \boldsymbol{\epsilon}_t.
\]

\noindent Here, $\mathbf{c}$ is a $K$-dimensional vector of intercept terms, $\Phi_1, \dots, \Phi_p$ are $K \times K$ matrices of autoregressive coefficients capturing the influence of past weeks’ case counts on the current week, and $\boldsymbol{\epsilon}_t \sim \mathcal{N}(\mathbf{0}, \Sigma)$ is a vector of random error terms with covariance matrix $\Sigma$. In our analysis, we set the lag order to $p = 25$ weeks. This choice was informed by exploratory correlation analyses, which indicated significant temporal dependencies between the jurisdictions’ weekly case counts extending up to 25 weeks. By including these lags, the VAR model is able to capture temporal dependencies in the spread of Mpox across jurisdictions.

\subsection{VAR-Lasso Regularization}

As the number of jurisdictions ($K$) and the lag order ($p$) increases, the number of parameters in a VAR model grows rapidly, risking overfitting. To address this, we employ Lasso, a regularization method that performs both variable shrinkage and selection. \citet{tibshirani1996regression} originally introduced Lasso for linear regression. Lasso introduces an L1 penalty on the model coefficients, which shrinks coefficients toward zero and sets some exactly to zero depending on the value of the penalty $\lambda$. This effectively performs variable selection while simultaneously estimating the model coefficients. In the context of VAR, this allows the model to identify which lags or past weeks’ case counts across the $K$ jurisdictions are most relevant for predicting current counts. Following \citet{hsu2008subset}, we define the lagged covariate vector for week $t$ as:
\[
\mathbf{X}_t = (1, \widetilde{\mathbf{y}_{t-1}}^\top, \widetilde{\mathbf{y}_{t-2}}^\top, \dots, \widetilde{\mathbf{y}_{t-p}}^\top)^\top.
\]
The design matrix is formed by arranging the vectors from all weeks as:
\[
\mathbf{X} = (\mathbf{X}_1, \dots, \mathbf{X}_n)^\top,
\]
which contains all lagged predictors corresponding to the responses in $\widetilde{\boldsymbol{Y}}$. Let 
\[
\boldsymbol{\beta} = \text{vec}(\mathbf{c}, \Phi_1, \dots, \Phi_p)
\]
denote the vector of coefficients in regression form. The Lasso estimator $\tilde{\boldsymbol{\beta}}$ is obtained by minimizing the penalized objective
\[
Q(\boldsymbol{\beta}) = (\widetilde{\mathbf{Y}} - \mathbf{X} \boldsymbol{\beta})^\top (\widetilde{\mathbf{Y}} - \mathbf{X} \boldsymbol{\beta}) + \lambda \sum_j |\beta_j|,
\] 
where $\lambda$ is the L1 penalty controlling the degree of shrinkage. When the L1 penalty $\lambda$ is small, the Lasso estimate reduces to the ordinary least squares (OLS) estimate. For larger values of $\lambda$, coefficients are shrunk toward zero, with some coefficients set exactly to zero, yielding a sparse model. In our analysis, Lasso regularization allows us to identify the most relevant past weeks’ values across the $K = 8$ jurisdictions while controlling for overfitting. The tuning parameter $\lambda$ is chosen using cross-validation, as suggested in \citet{tibshirani1996regression}.

\subsection{Model Fitting and Rolling Forecast Design} 

We implement a rolling forecast design for all 42 forecasted weeks between January and November 2024 for the eight selected jurisdictions. For each forecasted week, the VAR-Lasso model (lag = 25 weeks) is fit to all preceding data, starting from January 2023 up to two weeks before the forecasted week. The VAR-Lasso model is implemented using the BigVAR package in R \citep{BigVAR}. The modeling workflow is as follows:

\begin{enumerate}
    \item Apply a 4-week centered moving average to smooth the time-series and reduce noise
    \item Test each jurisdiction’s series for stationarity using the ADF test, and difference the series once
    \item Fit the VAR-Lasso model (lag = 25 weeks) using all preceding data up to two weeks before the forecasted week
    \item Generate two-week-ahead forecasts
    \item Convert the forecasts back to the original scale by reversing the differencing.
\end{enumerate}

\subsection{Benchmark models: AR-Lasso and Naive estimator}

To assess the predictive performance of the VAR–Lasso model, we compare its forecasts against two benchmark approaches: a univariate Auto-Regressive Lasso (AR–Lasso) model and a naïve moving-average estimator. The AR–Lasso model is a uni-variate autoregressive model with Lasso regularization that selects informative lag terms via cross-validation. The AR-Lasso model is implemented using the sparseVAR package in R \citep{sparseVAR_bigtime}. The naïve benchmark forecasts two weeks ahead by carrying forward the average number of cases observed over the preceding four weeks as the prediction; a similar comparator is used in the context of ILI forecasting in \citep{yang2017using}.

\subsection{Slope-weighted performance metrics}

Standard forecast metrics treat all time points equally when assessing prediction accuracy, without prioritizing properties that are epidemiologically relevant \citep{tabataba2017framework, armstrong2001evaluating}. In a public health context, features deemed more consequential may be weighted more heavily in forecast evaluation \citep{tabataba2017framework}. Periods of rising incidence are one such epidemiologically relevant period of an epidemic time-series and the urgency for public health action is often highest during such periods. To reflect this priority, we introduce slope-weighted error metrics that weight forecast errors by the magnitude of positive week-to-week growth, i.e. the magnitude of the positive slope in observed incidence. This approach ensures that periods of increasing transmission have proportionally greater influence on the evaluation, prioritizing model performance in the time windows where there is growth in cases. Below we formally define the slope-weighted metrics for assessing prediction accuracy. Let $\hat{y}_t$ denote the prediction for time $t$. The slope in observed incidence is given by

\begin{equation}
\Delta y^{\text{pos}}_t = \max(\widetilde{y_t} - \widetilde{y_{t-1}}, 0),
\label{eq1_delta}
\end{equation}
\vspace{5mm}
\noindent which captures only positive changes. The slope-weighted error metrics are as follows: 

\begin{itemize}
    \item The slope-weighted Root Mean Squared Error (RMSE) is  
    
    \begin{equation}
  \mathrm{Slope\text{-}weighted\ RMSE} = 
    \sqrt{\frac{\sum_{t} (\hat{y}_t - y_t)^2 \cdot \Delta y^{\text{pos}}_t}
    {\sum_{t} \Delta y^{\text{pos}}_t}}
    \label{eq2_rmse}
\end{equation}
    
    \item The slope-weighted mean absolute error (MAE) is
    
    \begin{equation}  \mathrm{Slope\text{-}weighted\ MAE} = 
    \frac{\sum_{t} |\hat{y}_t - y_t| \cdot \Delta y^{\text{pos}}_t}
    {\sum_{t} \Delta y^{\text{pos}}_t}
        \label{eq3_mae}
\end{equation}
    
    \item The slope-weighted bias is 
    
    \begin{equation}  \mathrm{Slope\text{-}weighted\ Bias} = 
    \frac{\sum_{t} (\hat{y}_t - y_t) \cdot \Delta y^{\text{pos}}_t}
    {\sum_{t} \Delta y^{\text{pos}}_t}
    \label{eq4_mae}
\end{equation}

\end{itemize}


\subsection{Phylogenetic analysis}
To characterize the evolutionary history of MPXV lineages introduced into SDC, we assembled and deduplicated alignments of MPXV genomes sampled in 2023 and 2024 and deposited in GenBank and/or GISAID (n=2571 genomes). Genomes were aligned using SQUIRREL, and a maximum likelihood tree was inferred using FASTTREE2 using a GTR+CAT20 substitution model \citep{price2010fasttree}. Distinct introductions into SDC, represented by 29 MPXV genomes, were identified using the migration model in TreeTime v0.11.4 \citep{sagulenko2018treetime}. We performed Bayesian phylogeographic inference in BEASTX v10.5.0 \citep{baele2025beast} on the five MPXV clusters, including at least one MPXV genome sampled in SDC. For each cluster, we ran two independent chains between 10 and 40 million generations under a GTR+F substitution model, strict molecular clock (prior: median $= 6 \times 10^{-5}$ substitutions/site/year; standard deviation $=2 \times 10^{-5}$) \citep{pekar2025transmission}, a SkyGrid coalescent prior \citep{baele2020hamiltonian}, and an asymmetric discrete traits model to infer ancestral geographic location throughout the phylogeny. Geographic locations prior to arrival in SDC were summarized from the posterior tree distribution using TreeSwift \citep{moshiri2020treeswift}.

\subsection{Code availability}
The code used in running the model, generating forecasts, evaluating the performance and creating visualizations is available at the public github repository: \\
\url{https://github.com/hanmacrad2/Mpox_VAR_Lasso_Automate}.

\newpage
\bibliography{References}

\subsection*{Author Contributions}
HC, JOW, MB, EAS, SS, RG, and NKM conceptualized the study. HC, JOW, RG, and NKM developed the statistical models. DV and RG assisted with the data. HC and JW carried out the formal analyses. HC, JOW, LR, RG, and NKM wrote the first draft of the manuscript. HC, JOW, EAS, MB, DV, RG, JW, LR, SS, RG and NKM reviewed and contributed to the final version.

\subsection*{Acknowledgements}
The authors gratefully acknowledge the authors from the originating laboratories and the submitting laboratories, who generated and shared through GISAID the viral genomic sequences and metadata on which this research is based (Supplementary Table).

\subsection*{Funding}
This project was made possible by cooperative agreement CDC-RFA-FT-23-0069 (Award: 1 NU38FT000006-01-00) from the CDC’s Center for Forecasting and Outbreak Analytics. Its contents are solely the responsibility of the authors and do not necessarily represent the official views of the Centers for Disease Control and Prevention.

\vspace{10cm}

\section*{Supplementary}

\renewcommand{\thetable}{S\arabic{table}}
\setcounter{table}{0}

\renewcommand{\thefigure}{S\arabic{figure}}
\setcounter{figure}{0}  

\subsection*{Prediction Interval Construction}

Two-step-ahead prediction intervals are constructed using a recursive forecasting framework applied to first-differenced time series. For the VAR-Lasso model, we fit the model to the differenced training data and obtain a one-step-ahead forecast for $t+1$ with prediction intervals using the \texttt{confint} option in the BigVAR package's \texttt{predict} function. We then extend the training series with the $t+1$ point forecast, refit the model, and generate a one-step-ahead forecast for $t+2$ with prediction intervals. The variance of the cumulative two-step change was calculated as
\[
\text{Var}(\Delta y_{t+1} + \Delta y_{t+2}) = \text{Var}(\Delta y_{t+1}) + \text{Var}(\Delta y_{t+2} \,|\, \widehat{y_{t+1}}).
\]
The variance calculation implicitly assumes independence of the forecast errors for time steps $t+1$ and $t+2$ conditional on $\widehat{y_{t+1}}$. This approach propagates uncertainty but may yield anti-conservative intervals, as it does not account for the covariance between errors across steps nor uncertainty arising from model selection. Prediction intervals from penalized regression methods such as Lasso do not capture uncertainty due to variable selection, shrinkage bias from regularization, or full parameter estimation \citep{BigVAR}.

For the AR-Lasso model, a similar two-step procedure was applied independently to each jurisdiction. At each forecast, we fit an AR model to the differenced series via maximum likelihood estimation using R's \texttt{arima} function, then generated direct two-step-ahead forecasts (\texttt{n.ahead = 2}) using the model's built-in \texttt{predict} function. The resulting standard errors inherently account for the correlation between one-step and two-step forecast errors through the autoregressive coefficients. These standard errors are used to construct 95\% prediction intervals, which are then converted from the differenced scale back to levels by cumulatively adding forecasted changes to the last observed value.

Future work could explore more rigorous methods for uncertainty quantification, including bootstrap resampling, de-biased Lasso, or Bayesian approaches, to obtain better-calibrated prediction intervals that account for the full uncertainty in penalized estimation procedures.

\begin{figure}[tbp!]
\centering
\includegraphics[width=1.0\textwidth]{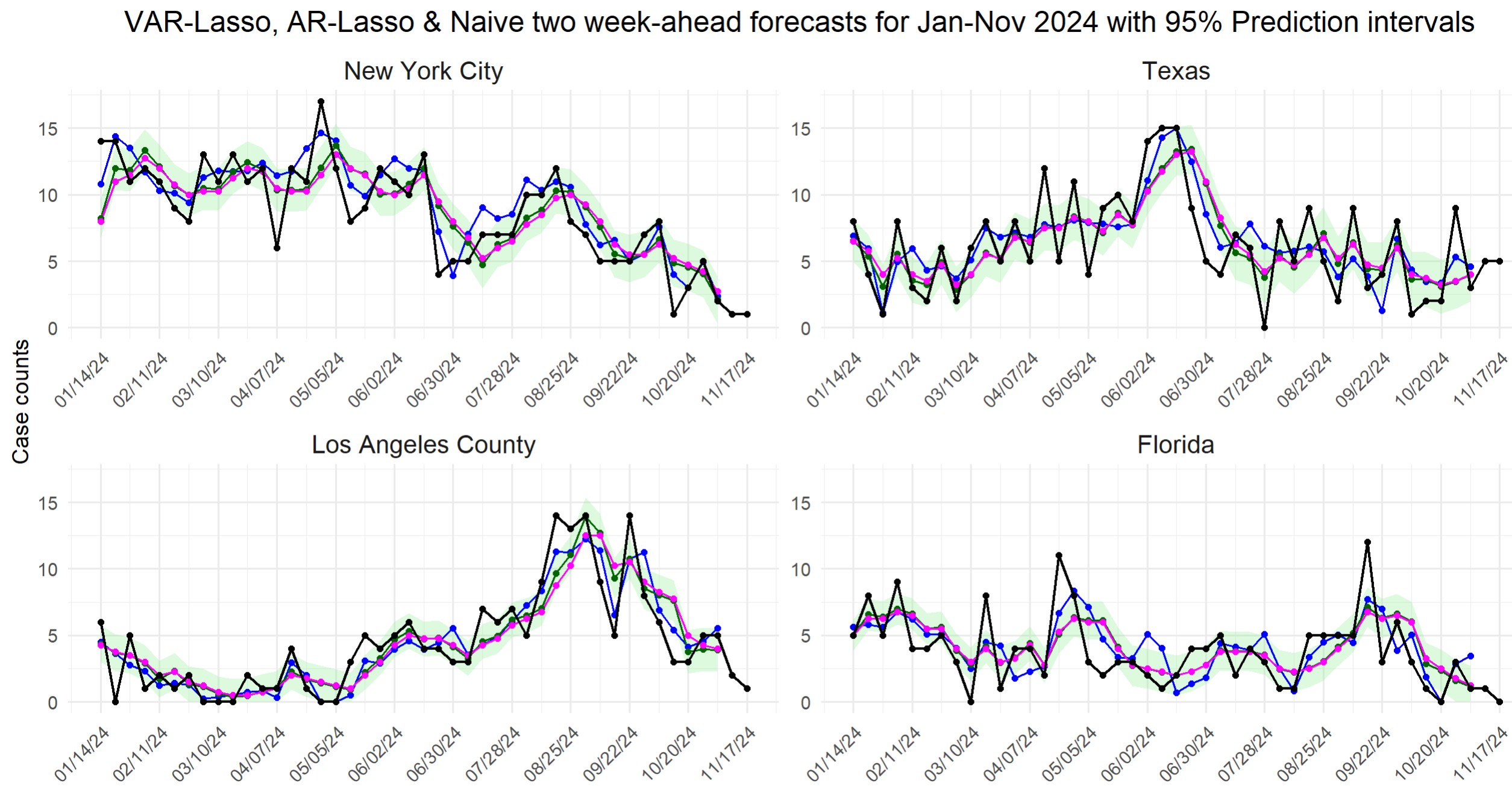}\\[-0.1em] 
\includegraphics[width=1.0\textwidth]{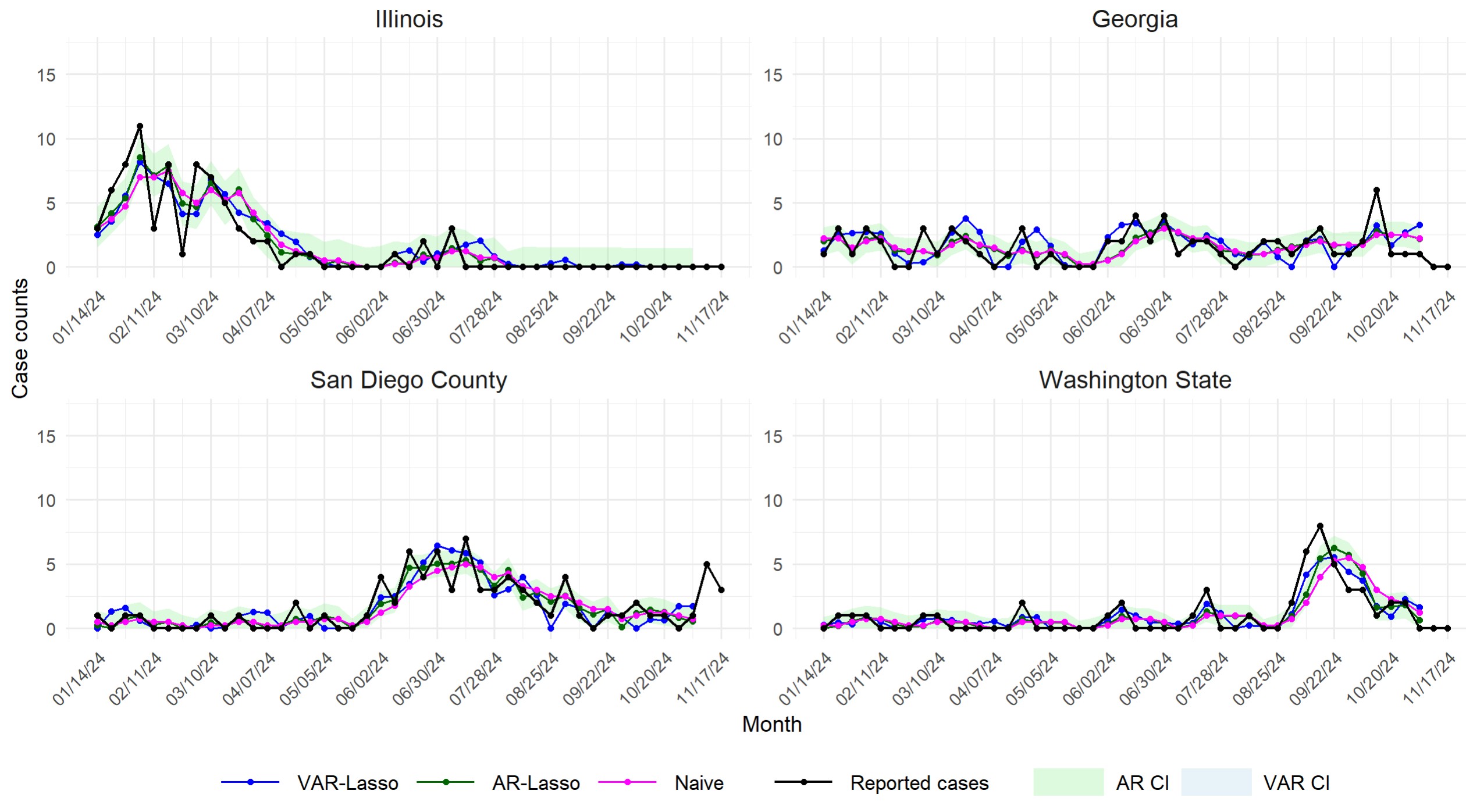}
\caption{\textbf{Forecasts with 95\% Prediction Intervals and weekly reported Mpox cases for the eight high-incidence US jurisdictions from January-November 2024}. Reported cases (black), VAR-Lasso (blue), VAR-Lasso prediction interval (light-blue), AR-Lasso (green), AR-Lasso prediction interval (light-green) and the naïve estimator (magenta).}
\label{fig:supp_7}
\end{figure}

\begin{table}[h]
\centering
\label{tab:clusters}
\begin{tabular}{ccccc}
\toprule
Cluster &
\makecell{Size \\ (genomes)} &
\makecell{San Diego \\ cases (\#)} &
\makecell{San Diego tMRCA: \\ median (95\% HPD)} &
\makecell{Location of origin \\ (posterior)} \\
\midrule
A & 433 & 4  & \makecell{2023.67 \\ (2023.57, 2023.69)} &
\makecell{CA (0.962) \\ CA-LAC (0.032)} \\

B & 142 & 19 & \makecell{2023.69 \\ (2023.57, 2023.75)} &
\makecell{CA (0.266) \\ CA-LAC (0.016) \\ IL (0.717)} \\

C & 244 & 4  & \makecell{2024.43 \\ (2024.38, 2024.47)} &
\makecell{CA (0.271) \\ CA-LAC (0.693) \\ NY-NYC (0.030)} \\

D & 102 & 1  & N/A &
CA (1.000) \\

E & 185 & 1  & N/A &
\makecell{Brazil (0.030) \\ Canada (0.721) \\ IL (0.022) \\ NY-NYC (0.175) \\ Netherlands (0.024) \\ Portugal (0.020)} \\
\bottomrule
\end{tabular}
\caption{ \textbf{Phylogenetic cluster characteristics for San Diego County mpox cases}. tMRCA, time of most recent common ancestor; HPD, highest posterior density; CA, California; IL, Illinois; CA-LAC, Los Angeles County; NY-NYC, New York City.}
\label{table:tab4}
\end{table}

\label{sec:supplementary}

\end{document}